\documentstyle[aps,amssymb,twocolumn,graphicx,epsfig]{revtex}

\def\be{\begin{equation}}
\def\ee{\end{equation}}
\def\bea{\begin{eqnarray}}
\def\eea{\end{eqnarray}}
\def\beq{\begin{eqnarray*}}
\def\eeq{\end{eqnarray*}}
\def\ba{\begin{array}}
\def\ea{\end{array}}

\def\v{{\bf v}}
\def\r{{\bf r}}

\begin{document}

\twocolumn[
{\Large\bf Brownian Motion of Grains and Negative Friction in Dusty Plasmas
}

\noindent
{\bf S.A.~Trigger}

\noindent
{\it Joint Institute for High Temperatures,
Russian Academy of Sciences, 
13/19 Izhorskaya Str., Moscow 127412, Russia;\\
Humboldt University, 110 Invalidenstr.,
D-10115 Berlin, Germany
}

\noindent
{\bf A.G.~Zagorodny}

\noindent
{\it Bogolyubov Institute  for Theoretical Physics, 
National Academy of Sciences of Ukraine\\
14B Metrolohichna Str., Kiev 03143, Ukraine\\
e-mail: azagorodny@gluk.org}

]

\begin{abstract}
Within the approximation of dominant charging collisions the explicit
microscopic calculations of the Fokker-Planck kinetic coefficients for
highly-charged grains moving in plasma are performed. It is shown that
due to ion absorption by grain the friction coefficient can be negative
and the appropriate threshold value of the grain charge is found. The
stationary solutions of the Fokker-Planck equation with the velocity
dependent kinetic coefficient are obtained and the considerable
deviation of such solutions from the Maxwellian distribution is
established.\\
PACS number(s) : 52.27.Lw, 52.20.Hv, 52.25.Fi
\end{abstract}

\vspace{0.5cm}

In the systems close to equilibrium Brownian particles keep stationary
random motion under the action of stochastic forces which are
compensated by the particle friction and thus, the work produced by the
Langevin sources is equal to the energy dissipated in course of the
Brownian particle motion. This energy balance is described by the
fluctuation-dissipation theorem in the form of the Einstein law.
Obviously, the fluctuation-dissipation theorem and the Einstein
relation can be violated in the case of non-equilibrium systems (even
in the stationary case), in particular in the open systems.  Starting
from the classical Lord Rayleigh work [1] many studies  of the
non-equilibrium motion of Brownian particles with
additional (inner, or external) energy supply have been performed. In
particular, such studies are of great importance
for physico-chemical [2, 3] and biological [4]
systems in which non-equilibrium Brownian particle motion is referred
as the motion of active Brownian particles.  Recently the dynamical and
energetic aspects of motion for the active Brownian particles have been
described on the basis of the Langevin equation and the appropriate
Fokker-Planck equation [5, 6].  The possibility of negative friction
(negative values of friction coefficient)  of the
Brownian particles was regarded, as a result of energy pumping. For
some phenomenological dependences of the friction coefficient as a
function of the grain's velocity the one-particle stationary
non-Maxwellian distribution function was found.

The traditional formulations of the non-equilibrium
Brownian motion are based on some phenomenological expressions for the
friction and diffusion coefficients. In particular, it means that
deviations from the Einstein relation, as well as the velocity
dependence of these coefficients are postulated and high level of
uncertainty for the application of such models to the real systems
takes place. Here we will consider another situation, when the
kinetic coefficients can be calculated exactly on the basis of the
microscopically derived Fokker-Planck  equation for dusty
plasmas [7,8].  It will be shown that in the case of strong
interaction parameter $\Gamma\equiv e^2 Z_g Z_i /aT_i \gg 1$ (here
$Z_g$, $Z_i$ are the charge numbers for the grains and ions
respectively, $a$ is the grain radius,  $T_i$  is the ion temperature)
the negative friction coefficient appears for some
velocity domain.  If the charging collisions are dominant, this
domain is determined by the inequalities:  $v^2<10 v^2_{\rm
Ti}(\Gamma-1)/(3\Gamma-1)$ for  $\Gamma\geq1$, but
$(\Gamma-1)\ll1$,$(v^2_{\rm Ti}\equiv T_i/m_i)$; and $v^2<2v^2_{\rm
Ti}\Gamma$ for $\Gamma\gg1$.  The physical  reason for manifestation of
negative friction is clear:  the cross-section for ion absorption by
grain increases with the relative velocity between the ion and grain
decrease due to the charge-dependent  part of the cross-section.
Therefore, for a moving highly-charged grain $(\Gamma\gg1)$ the
momentum transfer from ions to the grain in the direction of grain
velocity is higher than in the opposite direction.

Naturally, the Coulomb scattering and particle  friction related to the
ion-grain and  neutral-grain elastic collisions will increase the
threshold for negative friction and even can suppress it for some
plasma parameters.  However, the latter are rather sophisticated
processes, and their theoretical description within various
approximations can shadow the physics of the
phenomenon.  Some estimate of the influence of such
processes will be also done below, but the detailed description of
their influence should be a matter of further consideration.  On the
other hand, the problem of microscopic investigation of negative
friction in dusty plasmas  is so fundamental, that it deserves a
description of its simplest manifestation (which occurs in the case of
dominant charging collisions) to be considered in this Letter.

We start from the Fokker-Planck equation for the spherical grains in
dusty plasma with typical narrow charge distribution around a negative
value $q=e_eZ_g$, that permits to put
all grain charges to be equal. We also ignore the increase of the
grain mass [9, 10] assuming that neutral atoms generated in the course
of the surface electron-ion recombination escape from the grain surface
into a plasma. Than, for the conditions $l_i\gg \lambda_D$, $a$, where
$l_i$ is the ion mean free path length and $\lambda_D$ is  the plasma
screening (Debye) length, the friction and diffusion
coefficients in the Fokker-Planck kinetic equation $\beta(q,\v)$ and
$D(q,\v)$ [8] are given by:

\bea
\beta(q,\v)  &=& 
-\sum_{\alpha=e,i}{m_\alpha\over m_g}\!\int\!\!d\v'
{\v\cdot\v'\over v^2} \sigma_{\alpha g}(q,|\v-\v'|)|\v-\v'|
\nonumber \\
&\times&
 f_\alpha(\r,\v',t), \nonumber \\ 
D(q,\v) &=& \sum_{\alpha=e,i} \,{1\over2} 
\left({m_\alpha\over m_g}\right)^2 \int d\v' {(\v\cdot\v')^2\over v^2}\\
&\times&\sigma_{\alpha g}(q,|\v-\v'|)|\v-\v'| 
f_\alpha(\r,\v',t), \nonumber \\ 
\sigma_{g\alpha}(q,v)&=& \pi a^2 \left(1-{2e_\alpha 
q\over m_\alpha v^2a}\right) \theta
\left(1-{2e_\alpha q\over m_\alpha v^2 a}\right).  \nonumber 
\eea 
Here 
$\sigma_{g\alpha}(q,\v)$ is the cross-section for grain charging within 
the orbital motion limited (OML) theory. In this approximation all 
electrons and ions approaching to the grain on the distance less than 
$a$ are assumed to be absorbed. Subscript $\alpha=e,i$ labels plasma 
particle species. The rest of notation is traditional.   To include the 
processes of the electron, ion and atom scattering we have to summarize 
the appropriate  coefficients on the different type of the processes.  
In the case of dominant charging collisions integration in Eq.~(1) can 
be performed explicitly. The ion part of $\beta$ (which exceeds the 
electron one at least in $(m_iT_e/m_eT_i)^{1/2}$ times) is the 
following 

\bea 
\beta_i(q,\v) &=& -\sqrt{2\pi}{m_i\over m_g}a^2n_iv_{\rm 
Ti} \left[ I_1(\eta) +I_2(\eta,\Gamma)\right], \\ 
I_1(\eta)&=& 
\left(-{1\over2}+{1\over 4\eta}\right) \sqrt{{\pi\over\eta}} {\rm 
Erf}\sqrt{\eta} - {1\over2\sqrt{\eta}}  e^{-\eta},\nonumber \\ 
I_2(\eta, \Gamma) &=& {\Gamma\over\eta} \left[ {1\over2} 
\sqrt{{\pi\over\eta}} {\rm Erf} \sqrt{\eta}-e^{-\eta}\right].  
\nonumber 
\eea 
Here $n_a$  are the densities of  the electrons and 
ions, ${\rm Erf}\sqrt{\eta}$ is the error function and 
$\eta\equiv\eta(v)=v^2/2v^2_{\rm Ti}$. It is easy to see that the terms 
$I_1(\eta)$ and $I_2(\eta,\Gamma)$ describe the parts  of 
$\beta_i(q,\v)$ related to purely geometrical and charge-dependent 
collecting cross-sections, respectively.  The integration for 
$D_i(q,\v)$ leads to the expression:

\bea
D(q,\v) &=& {4\over3} \sqrt{2\pi} \left({m_i\over m_g}\right)a^2
n_iv_{\rm Ti} \left({T_i\over m_g}\right)\Bigl[K_1(\eta)
\\
&+&  K_2(\eta,\Gamma)\Bigr],\nonumber \\
K_1(\eta) &=& {3\over 16\eta} \left[ 2(\eta-1) e^{-\eta}\!\!+(2\eta^2
\!\!+\eta+1) \sqrt{{\pi\over\eta}} {\rm Erf}\sqrt{\eta}\right]\!\!,
\nonumber \\ 
K_2(\eta,\Gamma) &=& {3\Gamma\over 8\eta} (\eta+1) \left[ 
-2e^{-\eta} + \sqrt{{\pi\over \eta}} {\rm Erf}\sqrt{\eta}\right].  
\nonumber
\eea

As  follows from the physical reason and directly from Eq.~(3),
the coefficient $D(q, \v)$ is always positive. At the same time
for some values of $\eta$ and  $\Gamma$
the friction coefficient $\beta_i(q,\v)$ can be negative.
To find the root $\eta(\Gamma)$ of the equation
$\beta_i(\eta,\Gamma)=0$ let us consider  for two limiting cases
$\eta\ll1$ and $\eta\gg1$.  For $\eta\ll1$  Eq.~(2) gives

\be
\beta_i(\eta,\Gamma)=2A\left[ 1-\Gamma-{\eta\over5}(1-3\Gamma)\right],
\ee
where
\beq
A={1\over3}\sqrt{2\pi} \left({m_i\over m_g}\right)a^2n_iv_{\rm Ti}.
\eeq
Equation (4) has a  root  $\eta(\Gamma)$, which exists and is
small (according to the conditions of applicability for this expansion)
only for $\Gamma>1$, but $\Gamma-1\ll1$:
\be
\eta(\Gamma)=5{(\Gamma-1)\over 3\Gamma-1} \simeq {5\over2}(\Gamma-1).
\ee
Therefore the function $\beta_i(\eta,\Gamma)$ is negative at
$\eta<\eta(\Gamma)$, if $\Gamma>1$. Eq.~(4) shows
that for $\Gamma<1$, when $\beta_i$  is positive for all $\eta$, the
derivative $(d\beta_i(\eta,\Gamma)/d\eta)|_{\eta=0}$  changes its  sign
at $\Gamma=1/3$.  Near $\eta=0$ the
friction $\beta_i$ decreases as function $\eta$, if  $\Gamma<1/3$ and
increases, if $\Gamma>1/3$.

\begin{figure}[h]
\centering \epsfig{file=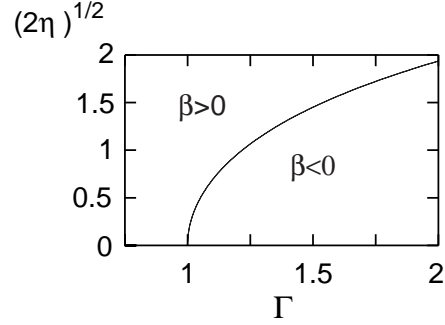,height=6cm}
 \caption{Numerical
solution of the equation $\beta(\eta,\Gamma)=0$ separating the
positive and negative values of the friction coefficient.}
\end{figure}

For  $\eta\gg 1$ and arbitrary values of $\Gamma$  Eq.~(2) gives

\be
\beta_i(\eta,\Gamma)={3\over2} A\sqrt{{\pi\over\eta}}\left[
1-{1+2\Gamma\over 2\eta}\right].
\ee
It means that for $\Gamma\gg1$ and large $\eta$ the equation
$\beta_i(\eta_1\Gamma)=0$ has  a root $\eta_1(\Gamma)\simeq
(1+2\Gamma)/2\sim\Gamma$.
Naturally, for all $\Gamma>1$ there exists appropriate
$\eta(\Gamma)$, which is the  root of that equation. This
conclusion is confirmed by the exact numerical calculations of
$\eta(\Gamma)$ (Fig.~1).  Asymptotically (for
$\eta\gg\max(1,\Gamma)$) $\beta_i(\eta,\Gamma)$ tends to zero as
$\sqrt{\eta^{-1}}$ and is positive for $\eta>\eta(\Gamma)$.  For the
case of the negative friction ($\Gamma >1$)  the maximum of the
coefficient $\beta_i(\eta,\Gamma)$ is located at the point
$\eta_m(\Gamma)\gg1$:

\be
\eta_m\simeq {3\over2}\left(1+2\Gamma\right),
\qquad \beta_i(\eta_m,\Gamma) \simeq A\sqrt{{2\pi\over 3(1+2\Gamma)}}.
\ee
For the diffusion coefficient $D_i(\eta,\Gamma)$ the expansions
for $\eta\ll1$ and $\eta\gg1$ lead to

\bea
D_i (\eta,\Gamma) &=&
4A \left({T_i \over m_g}\right) \left[1+{\Gamma\over2}
+{\eta\over10}(1+2\Gamma)\right], \, \, \eta\ll1 \\
D_i(\eta,\Gamma) &=& 4A\left({T_i\over m_g}\right) {3\over8}
(\pi\eta)^{1/2} \left[ 1+{1\over 2\eta} (1+2\Gamma)\right], \,
\eta\gg1.  \nonumber
\eea
The typical behaviour of $\beta_i(\eta,\Gamma)$ and $D_i(\eta,\Gamma)$
calculated numerically on the basis of Eqs.~(2), (3) is shown in
Figs.~2, 3.

\begin{figure}[h]
\centering \epsfig{file=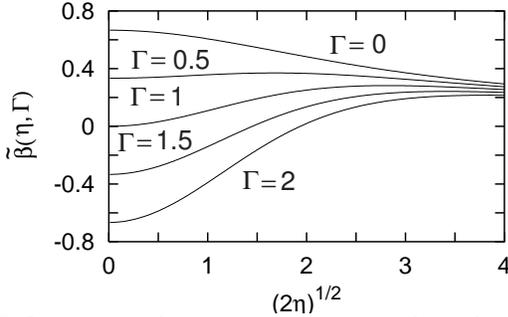,height=5cm} 
\caption{The
velocity dependences of the dimensionless friction coefficient
$\widetilde{\beta}(\eta,\Gamma)=I_1(\eta)+I_2(\eta,\Gamma)$ for
the different values of $\Gamma$. }
\end{figure}
\begin{figure}[h]
\centering \epsfig{file=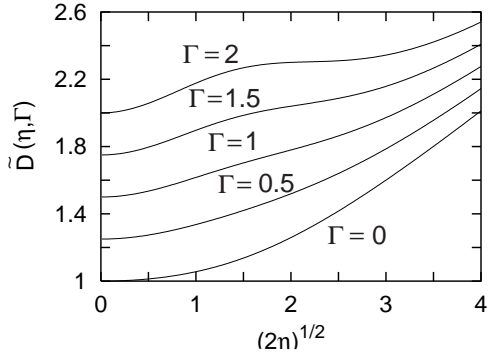,height=5cm} 
\caption{The same
dependences for the dimensionless diffusion coefficient
$\widetilde{D}(q,v)=K_1(\eta)+K_2(\eta,\Gamma)$.}
\end{figure}

The stationary solution of the Fokker-Planck equation with the kinetic
coefficients (2), (3) for the grain distribution function $f_g(q, v)$
is
\be f_g(q,v) ={C\over
D_i(q,v)} \exp\left[-\int\limits_0^v dv v{\beta_i(q,v)\over D_i(q,v)}
\right],
\ee
where $C$ is a constant, providing normalization $\int d\v f_g(q,v)=1$.
The velocity dependence of this solution for different values of
$\Gamma$ is shown in Fig.~4.

In order to get some analytical estimates let us consider
the vicinity of the point $\Gamma=1$. In such
case the  integration in Eq.~(9) leads to the non-Maxwellian
distribution function, which for $\Gamma>1$ possesses a maximum at
$v\neq0$:

\bea
f_g(q,\v)&=& \left({m_i\over 2\pi T_i}\right)^{3/2}
{\sqrt{\pi}\over2Y_{3/2}(1+\Gamma/2)} \Bigl[ 1  \\
&+&  {m_i v^2\over 2T_i^*}\, {1\over5}
(1+2\Gamma)\Bigr]^{-1} \exp\left\{ -{m_g v^2\over
2T_i^*} \Bigl[(1-\Gamma) \right. \nonumber \\
&+& \left.  {m_i v^2\over 2T_i^*}{1\over10}
\left(5\Gamma^2+4\Gamma-3\right)\Bigr]\right\}. \nonumber
\eea
Here

\bea
Y_\nu(\Gamma)&=& \int\limits_0^\infty d\eta \eta^{\nu-1}
\exp\left[-\alpha\eta(1-\Gamma)-\gamma\eta^2\right]\\
&=& (2\gamma)^{-{\nu\over2}}
{\sqrt{\pi}\over2} \exp\left[ {\alpha^2(1-\Gamma)^2\over
8\gamma}\right] D_{-\nu} \left({\alpha(1-\Gamma)\over
\sqrt{2\gamma}}\right) \nonumber \\[0.3cm]
T_i^* &=& 2T_i \left(1+{\Gamma\over2}\right),
\qquad \alpha={m_g\over 2m_i}
\left(1+{\Gamma\over2}\right), \nonumber \\
\gamma&=&{\alpha(5\Gamma^2+4\Gamma-3)\over 10(2+\Gamma)}
\nonumber
\eea
and $D_\nu$ is the cylindrical parabolic function.
Eq.~(10) is relevant, if the integral over $\eta$ converges, what
is valid for positive $\gamma(\Gamma>0,472)$. At the same time, as
was mentioned above,  for the applicability of the expansions (4), (8)
the inequality   $\Gamma<1+\delta$ with $\delta\ll1$ is required.
It is clear from Eqs.~(6), (8), that the  asymptotic
behaviour of the distribution function is non-exponential
$f_g(\eta)|_{\eta\rightarrow \infty} \sim \eta^{-(m_g/m_i)}$, but for
the values of $\Gamma$ under consideration  it is not essential for
calculations of the averages, due to the rapid convergence of the
integrals over $\eta$. In particular, the average kinetic energy
$K$ of grains is

\be
K(\Gamma)= \left({m_g\over
m_i}\right)T_i{Y_{5/2}(\Gamma)\over Y_{3/2}(\Gamma)}
\ee
that gives $K= 2,86 (m_g/m_i)^{1/2} T_i$ for $\Gamma=1$.
Thus, the ion absorption by grains can lead to grain heating and
grain average kinetic energy (or their
effective temperature) can be much higher, than the electron and ion
temperatures. This effect can be applied to the explanation of the
experimental data [11, 12], as it was already suggested in [7,8] on
the basis of the velocity independent approximation for $\beta$ and
$D$.

Let us consider now the case $\Gamma\gg1$.
In this case the  ratio $Q =
-(T_i/m_i)\beta(\eta,\Gamma)/D(\eta,\Gamma)$ can be
represented for all values of $\eta$ with a good accuracy by the
function $Q=(m_g/m_i)(\Gamma-\eta)/[(1+\Gamma)(\eta+\Gamma)]$ that
gives

\bea
f_g &=& {C\over D(\eta,\Gamma)} \exp
\left\{ {m_g\over m_i}\left[ {(3\Gamma+1)\over (2\Gamma-1)} \ln
{\Gamma(\eta+1)\over \Gamma+\eta} \right.\right.\\
&-& \left.\left. {1\over2}
\ln {\eta^2+(\Gamma+1)\eta+\Gamma\over \Gamma}\right]\right\},
\nonumber
\eea
where
\be
D_i(\eta, \Gamma)|_{\Gamma\gg1}\simeq {3\over2} A{T_i\over m_g}
\sqrt{\pi\eta} \left[
1+{\Gamma\over(\eta+{3\over4}\sqrt{\pi\eta})}\right].
\ee
It is evident, that  $f_g(\eta,\Gamma)$ given by Eq.~(13) is
non-exponential.

Finally for the domain  $\Gamma<0,472$ we can omit
the term $\sim v^4$  in Eq.~(10) and the distribution $f_g(q,v)$
becomes Maxwellian with the effective grain
temperature:
\be
T_{\rm eff} ={2T_i^*\over 1-\Gamma} ={2T_i(1+{\Gamma\over2})\over
1-\Gamma}.
\ee

The above conclusions are in a good agreement with the results of
numerical calculations of $f_g(\eta,\Gamma)$  on the basis of Eq.~(9)
(Fig.~4).

\begin{figure}[h]
\centering \epsfig{file=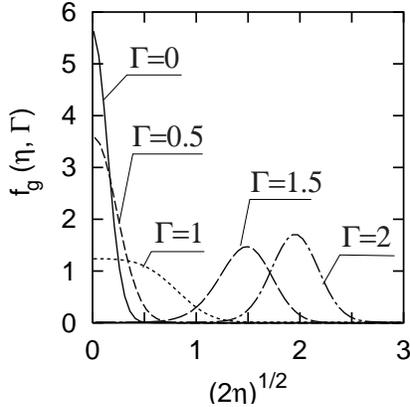,height=6cm} 
\caption{The same
dependences for the distribution function $f_g(\eta,\Gamma)$.}
\end{figure}

If we take into account the processes of atom-grain and ion-grain
elastic scattering  the coefficients $\beta(\eta,\Gamma)$ and
$D(\eta,\Gamma)$ should include additional terms.  To estimate, for
example, contribution of atom-grain and ion-grain scattering we can
consider $\beta(\eta)$ calculated  on the basis of the Eq.~(2) with the
appropriate transport cross-sections. For the case $\eta\ll1$ the
friction coefficient is

\bea
\beta(\eta,\Gamma) &=& 2A\left[
1-\Gamma+3{n_a\over n_i}\, \left({T_a m_a\over T_i m_i}\right)^{1/2}
\right.\nonumber \\
&-& \left. {\eta\over5} (1-3\Gamma) +6\Gamma^2\ln\Lambda\right].
\eea
The  negative friction exists for small $\eta$, if the Coulomb
scattering is strongly suppressed, when the Coulomb logarithm
$\ln\Lambda$ is small [13--15], that is typical for strong interaction.
The root of the function $\beta(\eta,\Gamma)$ is shifted to the region
of large, $\Gamma$, which is determined by the atom density $n_a$. The
result of rough estimate for the case $\lambda_{\rm Li}>\lambda_D>a$
($\lambda_{\rm Li}$ is the Landau length) gives

\be
\eta(\Gamma)={5\over 3\Gamma-1} \left[ \Gamma-1-3\sqrt{{Z_g \Gamma\over
\pi n_ia^3}} -{3n_a\over n_i} \, \left({T_a m_a\over T_i
m_i}\right)^{1/2} \right].
\ee
In general, a more detail consideration  of scattering processes and
the effects of strong interactions between the grains [15],  is
needed for the exact description of the region of negative friction.

In conclusion, ion absorption by grains can generate negative friction
and provide the substantial increase of the average grain kinetic
energy  in comparison with the temperatures of the other plasma
components.  Microscopical justification of negative friction on
kinetic level is presented and deviation  of the grain distribution
function from the Maxwellian distribution is found. For dusty plasma,
as for an open system, the fluctuation-dissipation  theorem in the
form of Einstein relation is not applicable.

\section*{Acknowledgment}
The authors thank  W.~Ebeling, U.~Erdmann, L.~Schimansky-Geier, and 
P.P.J.M.~Schram  for valuable discussions and the Netherlands 
Organization for Scientific Research (NWO) for the support of this 
work.  We also appreciate V.~Kubaichuk for the help in numerical 
calculations.


\begin{thebibliography}{99}
\bibitem{1}
J.W. Rayleigh. {\it The Theory of Sound}, vol. I, 2nd edition (Dover,
New-York) 1945.


\bibitem{2}
P-G. de Gennes. Rev. Mod. Phys. {\bf 57} 827 (1985).

\bibitem{3}
A.S. Mikhailov, and D. Meink\"ohn. In: Stochastic Dynamics, edited by
L. Schimansky- Geier, T. P\"oschel, vol. 484 of Lectures Notes in
Physics (Springer, Berlin, 1997) p. 334.

\bibitem{4} M. Scheinbein, and H. Gruler. Bull. Math. Biology {\bf 55}
585 (1993).

\bibitem{5}  F. Schweitzer, W. Ebeling, and B. Tilch. Phys. Rev. Lett. {\bf
80} 5044 (1998).

\bibitem{6}
U. Erdmann, W. Ebeling, L. Schimansky-Geier, and F. Schweitzer. Eur.
Phys. J.  {\bf B15} 105 (2000).

\bibitem{7}
A.G. Zagorodny, P.P.J.M. Schram, and S.A. Trigger. Phys. Rev. Lett.
{\bf 84} 3594 (2000).

\bibitem{8}
P.P.J.M. Schram, A.G. Sitenko, S.A. Trigger, and A.G. Zagorodny.
Phys. Rev. E {\bf 63} 016403 (2000).

\bibitem{9}
S.A. Trigger. Contrib. Plasma Phys. {\bf 41} 331 (2001).

\bibitem{10}
A.M. Ignatov, S.A. Trigger, W. Ebeling, and P.P.J.M. Schram. Phys.
Lett. {\bf A293} 141 (2001).

\bibitem{11}
A. Melzer, A. Homann, and A. Piel. Phys. Rev. E {\bf 53} 2757 (1996).

\bibitem{12}
H.M. Thomas, and G.E. Morfill. Nature
{\bf 379} 806 (1996).

\bibitem{13}
M.D. Kilgore, J.E. Daugherty, R.K. Porteous, and D.B. Graves, J. Apply
Phys.  {\bf 73} 7195 (1993).

\bibitem{14}
S.A. Khrapak, A.V. Ivlev, G.E. Morfill, and H.M. Thomas. {\it Ion 
drag force in complex plasmas}. In: 29 EPS Conference, abstract and 
report, Montreux, June 2002, to be published.

\bibitem{15}
S.A. Trigger, G.M.W. Kroesen, P.P.J.M. Schram, E. Stoffels, and
W.W. Stoffels. {\it Ion drag and Brownian motion in dusty plasmas}. In:  
29 EPS Conference, abstract and report, Montreux, June 2002, to be 
published.  
\end{thebibliography}
\end{document}